# Random Walk with Anti-Correlated Steps

John Noga[1]     Dirk Wagner[2]

## Abstract

We conjecture the expected value of random walks with anti-correlated steps to be exactly 1. We support this conjecture with 2 plausibility arguments and experimental data. The experimental analysis includes the computation of the expected values of random walks for steps up to 22. The result shows the expected value asymptotically converging to 1.

## 1 Introduction

Let $\{\mu_t\}$ be the horizontal random walk[3] that obeys the following rules:
1. The walk starts at $\mu_t = 0$
2. An urn contains $\delta$ white marbles and $\delta/2$ red marbles. At each step a marble is selected from that urn *without* replacement.
3. If the marble chosen is red $\mu_t$ increases by one ($\mu_{t+1} = \mu + 1$)
4. If the marble chosen is white $\mu_t$ decreases by one ($\mu_{t+1} = \mu -1$)
5. Let p=1/3 be the probability to draw a red marble. Accordingly, q=2/3 is the probability to draw a white marble.

n is the total number of marbles in that urn. Note that the absolute maximum positive is $\delta/2$, while the walk bounded by $-\delta$ in the negative direction. Note also the walk will terminate at $-\delta/2$.

This type of random walks involve anti-correlated steps, meaning the more red marbles are drawn (positive direction) the more likely are steps into the negative direction (white marbles are drawn). Anti-correlated walks are implemented simply be not replacing drawn marbles.

We are interested in the expected value $E[\max_t \mu_t]$ of $\{\mu_t\}$, that is for the expected value of a random walk with anti-correlated steps. [1] proves $E[\max_t \mu_t] = O(1)$. We are going to refine this result by showing that the expected value of random walks without replacement is likely to be exactly 1, thus $E[\max_t \mu_t] = 1$.

Applications for random walks with replacement can be found in caching algorithms. [1] introduces the RMARK algorithm – a random walk which cost is $E[\max_t \mu_t]$. [3] It can be shown that for random walks *with* replacement $E[\max_t \mu_t] = 1$.

---

[1] Department of Computer Science, California State University of Northridge, CA 91330. Email: jnoga@ecs.csun.edu

[2] Department of Computer Science, California State University of Northridge, CA 91330. Email: wagiboy@gmail.com

[3] Strictly put we consider random processes because the probability of steps changes with time.

## 2 Conjecture $E[\max_t \mu_t] = 1$

We conjecture $E[\max_t \mu_t] = 1$ for a random walk with anti-correlated step size. Support for this conjecture comes from two plausibility argument and our experimental data.

Plausibility argument one: Let $\{v\}$ be random walk with replacement with p=1/3, the probability to draw a red marble. Note that p remains constant during the walk. [4] shows the expected value of random walks with replacement is exactly 1, in short $E[\max_t v_t] = 1$.

For random walks without replacement p is changing, in fact $p_t$ is getting smaller as the walk progresses: At $p_0$=1/3. After an equal number of red and white marbles have been drawn $\mu_t$ returns to zero. At this point the probability that a red marble will be drawn next is no greater than 1/3. Let $p_1$ the probability to draw a red marble, then $p_1 < p_0$. The probability of subsequent returns of $\mu_t$ to zero exhibit an even further decrease such that $p_0 > p_1 > p_2 \llcorner$ . Note, if white marbles are drawn subsequent p do increase but at no effect to $E[\max_t \mu_t]$ because $\mu_t$ moves in the negative direction. Now when the probability p of drawing reds decreases over the course of a random walk, it is plausible that the expected value of random walks without replacement is not greater than for random walks with replacement, or $E[\max_t \mu_t] \le E[\max_t v_t] = 1$.

Plausibility argument two: As δ grows large drawing marbles does affect p very little. In fact, as $\delta \to \infty$ p does not change at all. In other words the random walk without replacement virtually becomes a random walk with replacement, thus $E[\max_t \mu_t] = E[\max_t v_t] = 1$.

## 3 Experimental Methods

In this section we provide experimental support for our conjecture $E[\max_t \mu_t] = 1$ by computing $E[\max_t \mu_t]$ for random walk of up to 30 steps. We implemented and compared several algorithms to efficiently compute the problem, including exhaustive computing, recursively selecting representatives, iteratively selecting representatives and linear optimizations.
As experimental method are implemented and compared on a 1.6 GHz Pentium M machine with 512 MB of main memory. The final experimental data has been computed on a 3.2??? GHz Pentium 4 with hyper threading and 2 GB??? of main memory.

### 3.1 Exhaustive Method

We used the following pseudo code to exhaustively compute $E[\max_t \mu_t]$.

```
input δ
n = 3/2* δ
total_max_µ=0;

for each of the n! permutations
  µ=0;
  max_µ=0;
  for each marble ∈ permutation
    if marble is red
      µ++;
      if( µ>max_µ ) max_µ=µ // find maximum µ
```

This method enabled us to comfortably compute $E[\max_t \mu_t]$ for the values δ=2,4,6,8. At δ=8 the program runtime on the (slower) 1.6 GHz Pentium M machine amounts to 3 ½ hours. As exhaustive computing represents only a crude first take on the problem no further comments need to be made.

## 3.2 Recursive Selection Representative Method

Computing $E[\max_t \mu_t]$ involves generating permutations of {$\mu_t$}, for example $\mu_t$=r$_1$, w$_1$,w$_2$,r$_2$,r$_3$,w$_3$,w$_4$,w$_5$. These permutations can be grouped by repeating subsequences of w's and r's, for example $\mu_t$=r w$^2$ r$^2$ w$^3$. Each group represents δ! (δ/2)! permutations of which each generates the same value $\mu_{t\ max}$. In other words the selecting one representative from group speeds up the computation of $E[\max_t \mu_t]$ by a factor of δ! (δ/2)!.

```
void Select_k_outof_n( int level, int loopstart )
{
    for( int i=loopstart; i<m_n-m_k+2+level; i++ )
    {
        m_selection[level]=i;

        if( level == m_finalLevel )
            SimulateRandomWalk( m_selection );
        else
            Select_k_outof_n( level+1, i+1 );
    }
}
```

Utilizing an iterative method we here able to push the envelope to δ=24 for computations lasting less than one hour.

## 3.4 Linear Optimizations

**End Walk at t=δ**

There is no need to let the algorithm simulate all drawings of $\mu_t$. At t=δ 3/2 of the walk has been completed. At this point there is no chance that $E[\max_t \mu_t]$ can grow, simply because there are not enough red marbles left to advance $\mu_t$ past 0 into the positive range. This implementation optimization speeds up the computation linearly by 1/3.

**Skip $E[\max_t \mu_t] = 0$ Selections**

The selection generating algorithm produces its sequences in lexicographic order. This fact enables us trim down the selections from the point forward when sufficiently white marbles have been drawn preventing $\mu_t$ to become positive. More precisely, the first red marble is drawn at t=δ/2 followed by a white marble. This implementation optimization speeds up the computation by ???.

## 4 Experimental Results

We have computed $E[\max_t \mu_t]$ as a function of the δ up to 30. As the graph in figure 1 shows $E[\max_t \mu_t]$ appears to be converging to the asymptotic boundary 1. Figure 1 also shows the graph of $E[\max_t \nu_t]$. Note that for δ>8 the 2 graphs are getting closer and closer, appearing both to be converging to 1. This converging behavior experimentally supports plausibility argument 2 that in the limit $E[\max_t \mu_t] = E[\max_t \nu_t] = 1$

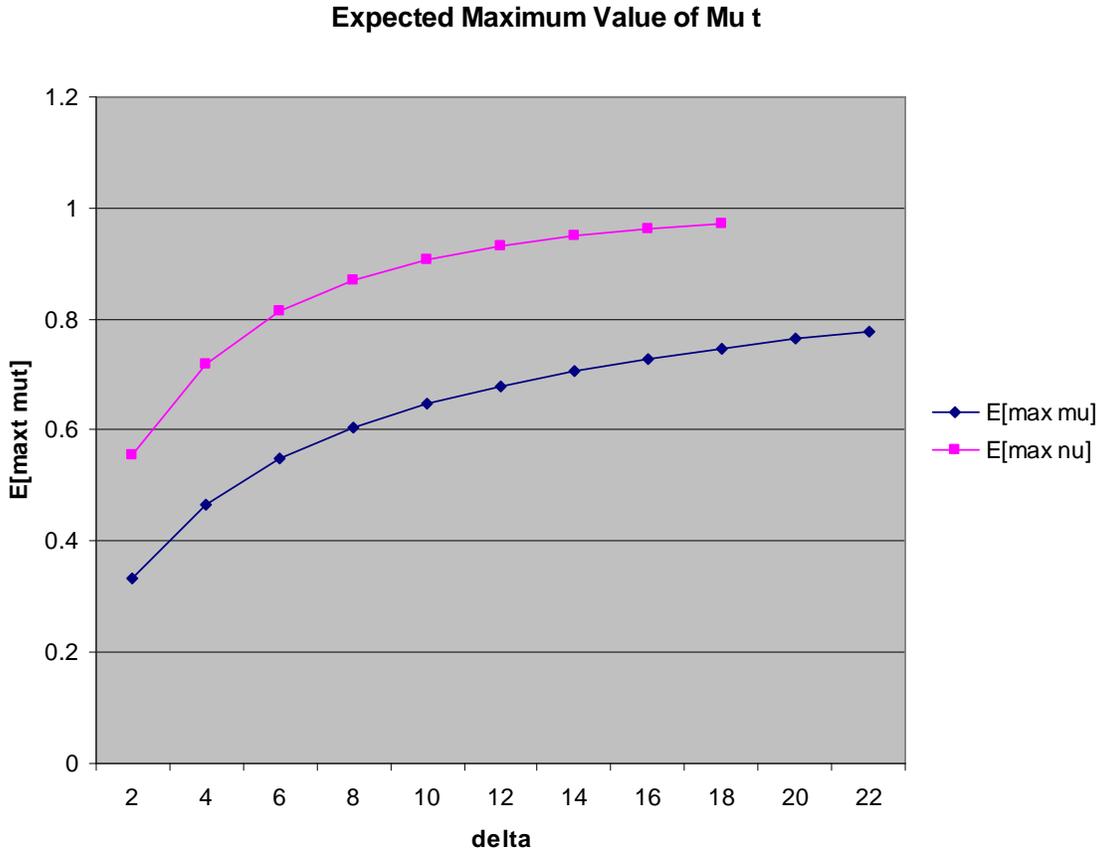

**Expected Maximum Value of Mu t**

Table 1 contains the numerical data used to create the graph of $E[\max_t \mu_t]$ in figure 1. It also contains the columns for the number of red and white marbles. The results of $E[\max_t \nu_t]$ are augmented with exact ratios are their respective run times.

| δ | #white | #red | Run time | $E[\max_t \mu_t]$ | $E[\max_t \mu_t]$ |
|---|---|---|---|---|---|
| 2 | 2 | 1 | 0.30 secs | 15/27 | 0.333333 |
| 4 | 4 | 2 | 0.30 secs | 524/729 | 0.466666 |
| 6 | 6 | 3 | 0.30 secs | 16017/19683 | 0.547619 |
| 8 | 8 | 4 | 0.30 secs | | 0.604040 |
| 10 | 10 | 5 | 0.30 secs | | 0.646354 |
| 12 | 12 | 6 | 1.26 secs | | 0.679595 |

| 14 | 14 | 7  | 3.64 secs              | 0.706570 |
| 16 | 16 | 8  | 21.66 secs             | 0.728995 |
| 18 | 18 | 9  | 169 secs (3 min)       | 0.747990 |
| 20 | 20 | 10 | 1342 secs (22 min)     | 0.764323 |
| 22 | 22 | 11 | 9301 secs (2 hrs 35 m) | 0.778541 |

Table 2 present data for $E[\max_t v_t]$

| δ  | #white | #red | Run time           | $E[\max_t v_t]$ | $E[\max_t v_t]$ |
|----|--------|------|--------------------|-----------------|-----------------|
| 2  | 2      | 1    | 0.30 secs          | 1/3             | 0.555556        |
| 4  | 4      | 2    | 0.30 secs          | 7/15            | 0.718793        |
| 6  | 6      | 3    | 0.30 secs          | 46/84           | 0.813748        |
| 8  | 8      | 4    | 0.30 secs          |                 | 0.868691        |
| 10 | 10     | 5    | 0.30 secs          |                 | 0.906453        |
| 12 | 12     | 6    | 1.26 secs          |                 | 0.931139        |
| 14 | 14     | 7    | 3.64 secs          |                 | 0.94919         |
| 16 | 16     | 8    | 21.66 secs         |                 | 0.961671        |
| 18 | 18     | 9    | 169 secs (3 min)   |                 | 0.971107        |
| 20 | 20     | 10   | 1342 secs (22 min) |                 |                 |